\documentstyle[aps,epsfig,multicol,eqsecnum]{revtex}

\newcommand{\s}[1]{~[#1]}
\newcommand{\zx}{\zeta(x_1)}

\newcommand{\ew}{\epsilon (\omega )}

\newcommand{\ts}{\theta_s}
\newcommand{\wc}{\frac{\omega^2}{c^2}}
\newcommand{\nn}{\nonumber}
\newcommand{\esw}{\epsilon_1(\omega )}
\newcommand{\eSw}{\epsilon_2(\omega )}

\newcommand{\iii}{\int^{\infty}_{-\infty}}

\newcommand{\bqe}{\begin{eqnarray}}
\newcommand{\eqe}{\end{eqnarray}}

\newcommand{\fr}{\frac{1}{2}}
\newcommand{\sfr}{^{\frac{1}{2}}}

\newcommand{\w}{\omega }

\newcommand{\aop}{\alpha_0(p, \omega )}
\newcommand{\aoq}{\alpha_0(q, \omega )}
\newcommand{\alp}{\alpha  (p , \omega )}
\newcommand{\aq}{\alpha (q, \omega )}

\newcommand{\la}{\langle}
\newcommand{\ra}{\rangle}

\newcommand{\fwc}{\frac{\omega}{c}}

\newcommand{\e}{\epsilon}

\newcounter{tr}

\newcommand{\sctr}[1]{\setcounter{tr}{#1}}
\newcommand{\sceq}{\addtocounter{equation}{-1}}

\begin{document}

\title{The Design of Random Surfaces with Specified
        Scattering Properties: \\Surfaces that Suppress
        Leakage}

\author{A. A. Maradudin$^a$, I. Simonsen$^{a,b}$, T. A. Leskova$^{c}$,
         E. R. M\'endez$^d$}  

\address{
$^a$Department of Physics and Astronomy
    and Institute for Surface and Interface Science,\\
    University of California,
    Irvine, CA 92697, U.S.A.\\
$^b$Department of Physics,
         The Norwegian University of Science and Technology,\\
         N-7491 Trondheim, Norway\\
$^c$Institute of Spectroscopy, Russian Academy of Sciences, 
         Troitsk 142092, Russia\\
$^d$Divisi\'on de F\'{\i}sica Aplicada, 
      Centro de Investigaci\'on Cient\'{\i}fica y de Educaci\'on
      Superior de Ensenada,\\
      Apartado Postal 2732,
      Ensenada, Baja California 22800, M\'exico 
}

\maketitle

\begin{abstract} 
  We present a method for generating a one-dimensional random metal
  surface of finite length $L$ that suppresses leakage, i.e. the
  roughness-induced conversion of a surface plasmon polariton
  propagating on it into volume electromagnetic waves in the vacuum
  above the surface.  Perturbative and numerical simulation
  calculations carried out for surfaces generated in this way show
  that they indeed suppress leakage.
\end{abstract}

\section{Introduction}

The great majority of theoretical studies of scattering from randomly 
rough surfaces have been devoted to the direct problem, in which the 
surface profile function and its statistical properties are 
specified, and it is the angular distribution of the scattered 
intensity, and its polarization properties, that are sought. 
Comparatively little attention has been devoted to the inverse 
problem.  In the usual formulation of this problem scattering data 
are provided by experimentalists, and the surface profile,  or some 
statistical properties of it, are sought.  This kind of  inverse 
problem can be called a passive inverse problem, because the theorist 
has no control over the experimental data provided.  In contrast, in 
the present work we consider a different type of inverse problem, one 
that we can characterize as an active inverse problem.  In this type 
of problem, the angular distribution of the scattered intensity is 
specified, and the problem is to design, and ultimately to fabricate, 
a surface that generates that angular distribution.

This active type of inverse problem has been studied even less than 
the passive type, yet it arises in a variety of contexts of both a 
technical and a basic physics nature.  In this paper, we present an 
illustrative example of the design of a random surface with specified 
scattering properties that is prompted by the observation that as a 
surface plasmon polariton propagates across a randomly rough metal 
surface it continuously loses energy through its roughness-induced 
conversion into volume electromagnetic waves in the vacuum above the 
random surface, that propagate away from the surface.  This leakage, 
as it is called, interferes with the determination of the Anderson 
localization length of the surface plasmon polariton by means of 
numerical simulation calculations, or experimental measurements, of 
its transmissivity as a function of the length of the random segment 
of the surface\s{1-3}.  The use of a random surface in such studies 
that suppresses leakage therefore facilitates the investigation of the 
strong localization of surface plasmon polaritons by random surface 
roughness.

In the approach to the suppression of leakage taken by Sornette and 
his colleagues\s{1,2}, it was assumed that the random surface was not 
planar on average, but periodic, so that the dispersion curve of the 
surface plasmon polaritons supported by the mean surface displays a 
gap at the boundary of the one-dimensional first Brillouin zone 
defined by the period of the mean surface.  Leakage should then 
either vanish or substantially decrease for the surface plasmon 
frequency at the band edge.  However, this was not observed in the 
numerical simulation calculations of leakage carried out in Refs. 3 
and 4.

In this work we present an approach to designing a one-dimensional 
random surface that suppresses the leakage of a surface plasmon 
polariton as it propagates across it that differs from that proposed 
by Sornette {\it et al.}\s{1,2}.  Although the power spectrum of the 
resulting surface is nonzero in a narrow range of wave numbers, that 
surface is not periodic on average.  However, as with the surface 
proposed by Sornette {\it et al.}, our surface is specific to the 
frequency of the surface plasmon polariton propagating across it:  if 
that frequency is changed, a new surface has to be designed.

\begin{figure}[t]
    \begin{center}
        \epsfig{file=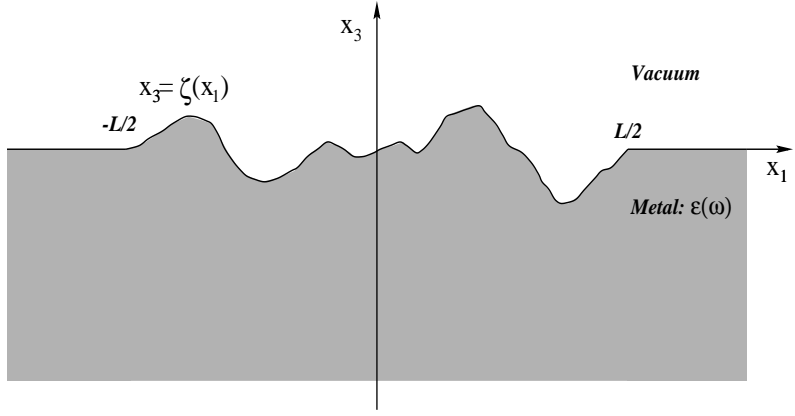,width=10.5cm,height=5.5cm} 
    \end{center}
    \caption{The system studied in this paper.}
\end{figure}
\vspace*{0.8cm}

\section{The Scattered field}

We study the scattering of a p-polarized surface plasmon polariton of 
frequency $\w$ propagating in the  $x_1$-direction that is incident 
on a segment of a one-dimensional randomly rough surface defined by 
the equation $x_3 = \zx$.  The surface profile function $\zx$ is 
assumed to be a single-valued function of $x_1$ that is nonzero only 
in the interval $-L/2 < x_1 < L/2$ (Fig. 1).  The region $x_3 > \zx$ 
is vacuum; the region $x_3 < \zx$  is a metal characterized by an 
isotropic, frequency-dependent, complex dielectric function $\ew = 
\e_1(\w ) +i\e_2(\w )$.  We are interested in the frequency range in 
which $\e_1(\w ) < -1, \e_2 (\w ) > 0$, within which surface plasmon 
polaritons exist.  We assume that the surface roughness is 
sufficiently weak that the surface profile function $\zx$ satisfies 
the conditions for the validity of the Rayleigh hypothesis\s{5}.  In 
this case the single nonzero component of the magnetic field in the 
vacuum region $x_3 > \zx_{max}$ can be written as the sum of the 
fields of the incident and scattered waves
\bqe
H^>_2(x_1,x_3|\w ) &=& \exp [ik(\w )x_1-\beta_0(\w ) x_3 ]
   +  \iii \frac{dq}{2\pi} R^>(q,\w )\exp [iqx_1+i\aoq x_3] ,
\eqe
while in the region of the metal , $x_3 < \zx_{min}$,
\bqe
H^<_2(x_1,x_3|\w ) &=& \exp [ik(\w )x_1+\beta (\w ) x_3 ]
   + \iii \frac{dq}{2\pi} R^<(q,\w )\exp [iqx_1-i\aq x_3] .
\eqe
In Eqs. (2.1)-(2.2) $k(\w )$ is the surface plasmon polariton wave number,
\bqe
k(\w ) = \fwc \left[ \frac{\ew}{\ew + 1}\right]\sfr = k_1(\w ) + ik_2(\w ),
\eqe
with
\sctr{1}
\bqe
k_1(\w ) &=& \fwc \left( \frac{|\esw |}{|\esw |-1} ]\right)\sfr > 0\\ 
\sctr{2}\sceq
k_2(\w ) &=& \fr \fwc \left( \frac{|\esw |}{|\esw |-1}\right)\sfr 
\frac{\eSw}{|\esw |(|\esw |-1)} > 0,
\eqe
while the functions
\sctr{1}
\bqe
\beta_0(\w ) &=& \fwc \left[ \frac{-1}{\ew + 1}\right]\sfr\\ \sctr{2}\sceq
\beta (\w ) &=& -\ew\fwc \left[ \frac{-1}{\ew +1}\right]\sfr
\eqe
characterize the exponential decay of the field of the surface 
plasmon polariton with increasing distance from the interface into 
the vacuum and the metal, respectively.  The functions $R^>(q,\w )$ 
and $R^<(q,\w )$ are the scattering amplitudes of the surface plasmon 
polariton in the vacuum and in the metal, respectively, and
\sctr{1}
\bqe
\aoq &=& \left( \wc - q^2\right)\sfr \hspace*{.5in} Re\aoq > 0, 
Im\aoq > 0\\ \sctr{2}\sceq
\aq &=& \left( \ew \wc - q^2\right)\sfr \hspace*{.2in} Re\aq > 0, Im\aq > 0.
\eqe
The scattering amplitude $R^>(q,\w )$ satisfies the reduced Rayleigh 
equation\s{6}
\sctr{0}
\bqe
R^>(p,\w ) &=& - \frac{\ew - 1}{\ew\aop + \alp} \left\{ \frac{J(\alp - 
i\beta_0(\w )|p-k (\w ))}{\alp - i\beta_0(\w )} \nn\right. \\
&& \hspace*{2in} \times [pk(\w ) + i\alp \beta_0 (\w )]\nn\\
&& + \iii \left. \frac{dq}{2\pi} \frac{J(\alp - \aoq |p-q)}{\alp - 
\aoq} [pq + \alp \aoq ]R^>(q,\w )\right\} ,\hphantom{dkd}
\eqe
where
\bqe
J(\gamma |Q) = \iii dx_1 e^{-iQx_1} (e^{-i\gamma\zx} -1) .
\eqe

This equation will be solved perturbatively and numerically.

\setcounter{equation}{0}
\section{The Total Scattered Power}

From Eq. (2.1) we see that the scattered field in the 
vacuum region can be written in the form
\bqe
H^>_2(x_1,x_3|\w )_{sc} = \iii \frac{dq}{2\pi} R^>(q,\w ) 
e^{iqx_1+i\aoq x_3}.
\eqe

The total scattered power, normalized by the total power in the 
incident surface plasmon polariton, is
\bqe
S(\w ) = \frac{P_{sc}}{P_{inc}} ,
\eqe
where
\sctr{1}
\bqe
P_{sc} &=& L_2 \frac{\w}{16\pi^2} 
\int^{\frac{\pi}{2}}_{-\frac{\pi}{2}} d\ts \cos^2\ts \left|R^>\left( 
\fwc \sin\ts ,\w \right)\right|^2 \\ \sctr{2}\sceq
P_{inc} &=& L_2 \frac{c^2}{16\pi\w} \frac{\e^2(\w ) - 1}{(-\ew )^{3/2}} ,
\eqe
with $L_2$ the length of the surface along the $x_2$-axis.  The 
scattering angle $\ts$, measured clockwise from the $x_3$-axis, is 
related to the wavenumber $q$ by $q = (\w /c)\sin\ts$.  We therefore 
find that
\sctr{0}
\bqe
S(\w ) = \frac{1}{\pi} \wc \frac{(-\ew )^{3/2}}{\e^2(\w ) - 1} 
\int^{\frac{\pi}{2}}_{-\frac{\pi}{2}} d\ts \cos^2\ts \left| R^> 
\left(\fwc \sin\ts ,\w \right) \right|^2 .
\eqe

Since the integrand in Eq. (3.3a) (and in Eq. (3.4)) is non-negative, 
we see that the only way in which leakage can be suppressed, i.e. the 
only way in which $P_{sc}$ can be made to vanish, is to design a 
one-dimensional  random surface for which the amplitude $R^>(q,\w )$ 
is identically zero for $-(\w /c) < q < (\w /c)$.  In the next 
section we introduce one such surface.

\setcounter{equation}{0}
\section{The Random Surface}

We write the surface profile function $\zx$ in the form
\bqe
\zx = T(x_1)s(x_1) ,
\eqe
where $s(x_1)$ is a single-valued function of $x_1$ that is 
differentiable and constitutes a stationary, zero-mean, Gaussian 
random process defined by
\sctr{1}
\bqe
\la  s(x_1)\ra &=& 0\\ \sctr{2}\sceq
\la s(x_1)s(x'_1)\ra &=& \delta^2W(|x_1-x'_1|)\ra \\ \sctr{3}\sceq
\la s^2(x_1)\ra &=& \delta^2,
\eqe
where the angle brackets denote an average over the ensemble of 
realizations of $s(x_1)$.  The function $T(x_1)$ serves to restrict 
the nonzero values of $s(x_1)$ to the interval $-L/2 < x_1 < L/2$. 
One form $T(x_1)$ can have is
\sctr{0}
\bqe
T(x_1) = \theta (\frac{L}{2} + x_1) \theta (\frac{L}{2} - x_1),
\eqe
where $\theta (x_1)$ is the Heaviside unit step function.  
A  smoother, differentiable version of $T(x_1)$ is provided by
\bqe
T(x_1) = \frac{1 + \cosh \fr \beta L}{\cosh \beta x_1 + \cosh \fr \beta L} ,
\eqe
where the parameter $\beta$ controls the range of $x_1$  values over 
which $T(x_1)$ decreases from 1 to 0.  A value of $\beta$ given by 
$100/L$ cuts off $s(x_1)$ smoothly.

The power spectrum of $s(x_1), g(|Q|)$, is defined by
\bqe
g(|Q|) = \iii dx_1 e^{-iQx_1}W(|x_1|) .
\eqe
A surface that suppresses leakage is defined by the power spectrum (Fig. 2)
\bqe
 \label{power}
g(|Q|)  = \frac{\pi}{2\Delta k} [\theta (Q-k_{min})\theta (k_{max}-Q) 
+ \theta (-Q-k_{min})\theta (k_{max}+Q)] ,
\eqe
where
\sctr{1}
\bqe
k_{min} &=& 2k_1(\w ) - \Delta k\\ \sctr{2}\sceq
k_{max} &=& 2k_1(\w ) + \Delta k ,
\eqe
and $\Delta k$ must satisfy the inequality
\sctr{0}
\bqe
\Delta k < k_1(\w ) - (\w /c) .
\eqe

\begin{figure}[t]
    \begin{center}
        \epsfig{file=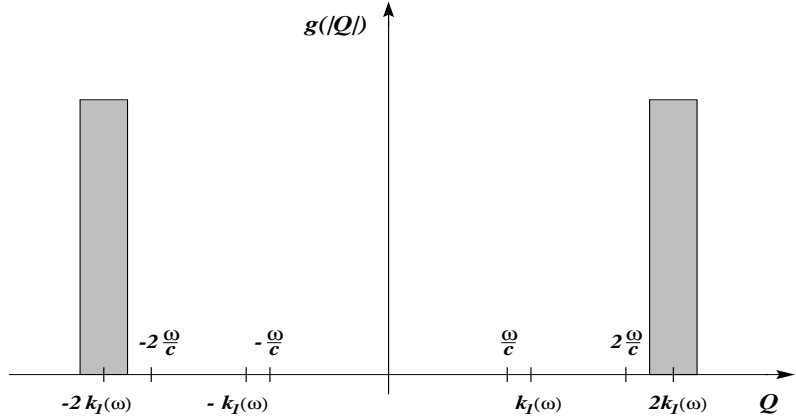,width=10.5cm,height=5.5cm} 
    \end{center}
    \caption{The power spectrum of the surface roughness assumed in this work}
\end{figure}
\vspace*{0.8cm}

That a surface characterized by the power spectrum (\ref{power})
suppresses leakage can be seen from the following argument.  The
incident surface plasmon polariton has a wave number whose real part
is $k_1(\w )$.  After its first interaction with the surface roughness
the real part of its wave number will lie in the two intervals
$(3k_1(\w ) - \Delta k, 3k_1(\w ) + \Delta k)$ and $(-k_1(\w )-\Delta
k,-k_1(\w )+\Delta k)$.  This is because the wave numbers in the
spectrum of the surface roughness with which $k_1(\w )$ can combine
lie in the intervals $(2k_1(\w )-\Delta k, 2k_1(\w ) + \Delta k)$ and
$(-2k_1(\w )-\Delta k, -2k_1(\w ) + \Delta k)$.  For the same reason,
after its second interaction with the surface roughness the real part
of the wave number of the surface plasmon polariton will lie in the
three intervals $(5k_1(\w ) - 2\Delta k, 5k_1(\w ) +2\Delta k)$,
$(k_1(\w ) - 2\Delta k, k_1(\w ) + 2\Delta k)$, and $(-3k_1(\w
)-2\Delta k, -3k_1(\w ) + 2\Delta k)$.  After three interactions with
the surface roughness the real part of its wave number will lie in the
four intervals $(7k_1(\w )-3\Delta k, 7k_1(\w ) + 3\Delta k)$,
$(3k_1(\w )-3\Delta k, 3k_1(\w )+3\Delta k)$, $(-k_1(\w ) - 3\Delta k,
-k_1(\w ) + 3\Delta k)$, and $(-5k_1(\w ) - 3\Delta k,-5k_1(\w
)+3\Delta k)$, and so on.  Thus, for example, if $-k_1(\w )+3\Delta k
< -(\w /c)$, so that $\Delta k < \frac{1}{3} (k_1(\w ) - (\w /c))$,
after three scattering processes the surface plasmon polariton will
not have been converted into volume electromagnetic waves.  In
general, if we wish the surface plasmon polariton to scatter $n$ times
from the surface roughness without being converted into volume
electromagnetic waves, we must require that $\Delta k < \frac{1}{n}
(k_1(\w ) - (\w /c))$.

\vspace*{0.8cm}
\begin{figure}[b]
    \begin{center}          
        \epsfig{file=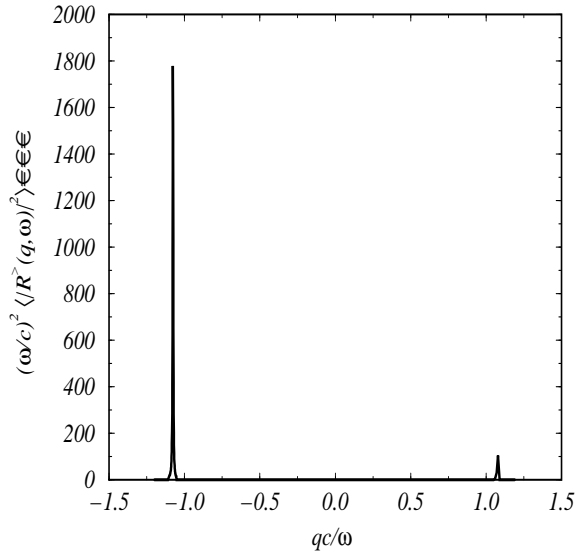,width=7.5cm,height=7.5cm} 
    \end{center}
 \caption{A perturbative result for $(\w /c)^2 \la |R^>(q,\w 
   )|^2\ra$ as a function of $(cq/\w )$ for a surface plasmon
   polariton, whose frequency corresponds to a vacuum wavelength
   $\lambda = 457.9 $ nm, propagating across a random silver surface
   $(\ew = - 7.5 + i0.24)$ of length $L = 20 \lambda$, defined by the
   power spectrum (4.6) with $\Delta k = 0.3 (k_1(\w ) - (\w /c))$,
   $\delta = 10$ nm, and Eq.  (4.3).}
\end{figure}

\vspace*{0.8cm}
\begin{figure}[b]
    \begin{center}          
        \epsfig{file=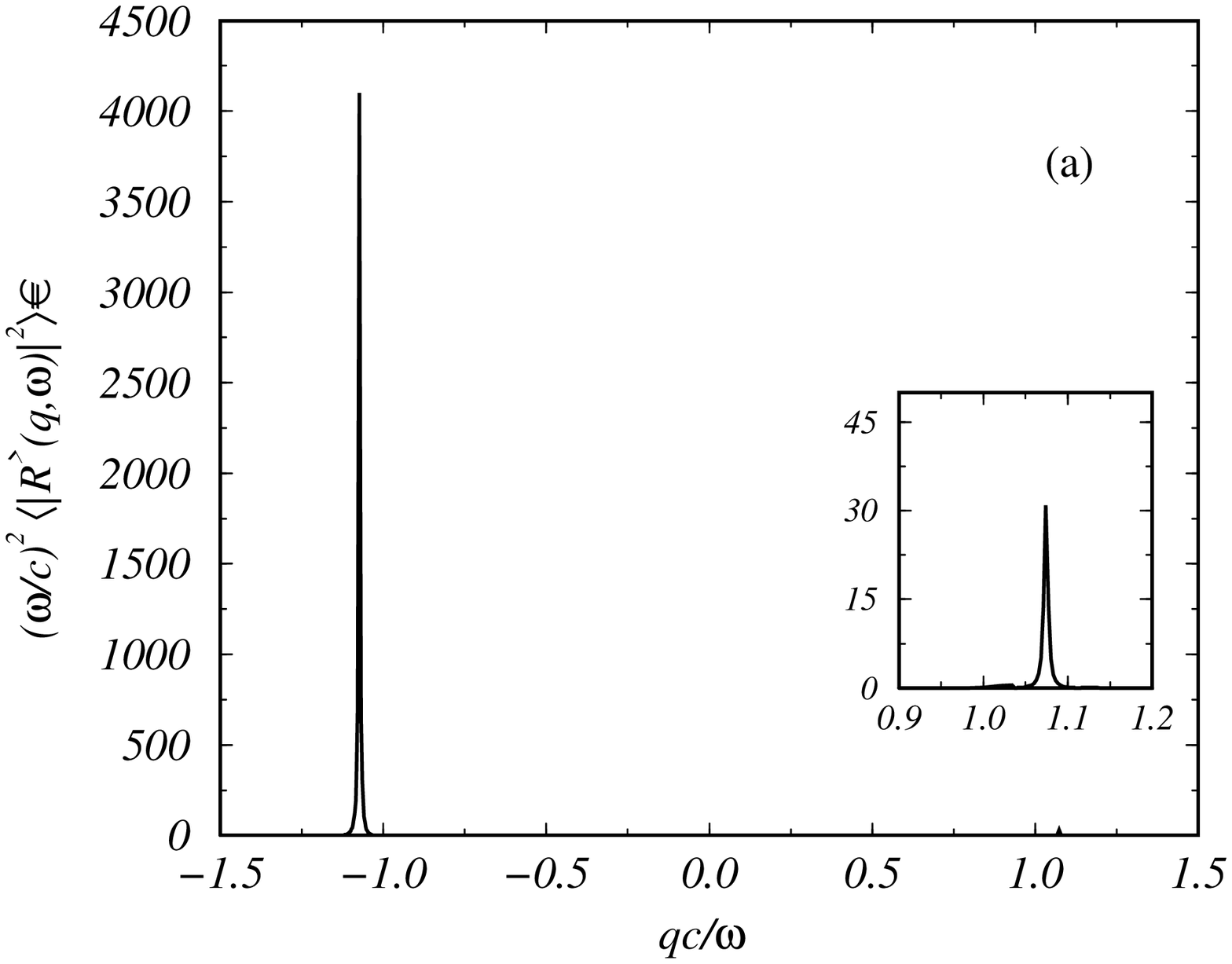,width=7.5cm,height=7.5cm} \hspace*{0.9cm}
         \epsfig{file=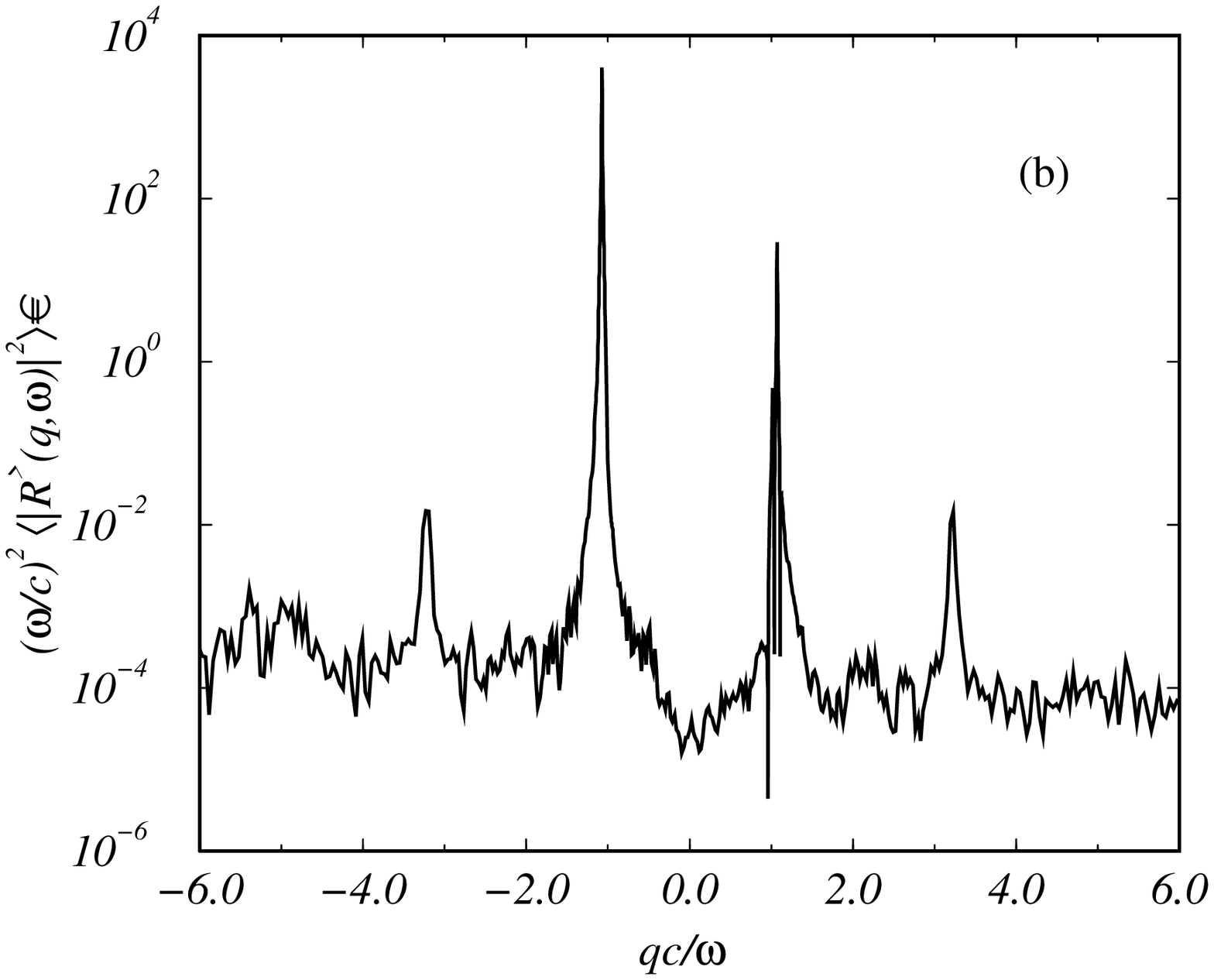,width=7.5cm,height=7.5cm}
    \end{center}
 \caption{A computer simulation result for $(\w /c)^2 \la |R^>(q,\w 
   )|^2\ra$ as a function of $(cq/\w )$, plotted on (a) a linear-linear
   scale and (b) a linear-log scale, for a surface plasmon polariton,
   whose frequency corresponds to a vacuum wavelength $\lambda =
   457.9$ nm, propagating across a random silver surface $(\ew = - 7.5
   + i0.24)$ of length $L = 20 \lambda$, defined by the power spectrum
   (4.6) with $\Delta k = 0.3 (k_1(\w ) - (\w /c))$, $\delta = 30$ nm,
   and Eq.  (4.4). The results for 50 realizations of the surface
   profile function were averaged to obtain the results plotted in
   this figure.}
\end{figure}

To show that this approach to suppressing leakage works, in Fig.~3 we 
present a plot of $(\w /c)^2\la |R^>(q,\w )|^2\ra$ as a function of 
$cq/\w$ for a silver surface characterized by the power spectrum 
(4.6) with $\Delta k = 0.3 (k_1(\w )-(\w  /c))$ and $\delta = 10 $ 
nm.  The frequency of the surface plasmon polariton corresponds to a 
vacuum wavelength  of $\lambda = 457.9$ nm, and the dielectric 
function of silver at this frequency is $\ew = -7.5 + i0.24$.  The 
calculation of $\la |R^>(q,\w )|^2\ra$ was carried out as an 
expansion in powers of the surface profile function through terms of 
fourth order.    For this $R^>(q,\w )$ was calculated as an expansion 
in powers of $\zx$ through terms of third order (there is no zero 
order term).  The expression (4.3) was used for $T(x_1)$, and the 
length of the random segment of the metal surface was $L = 
20\lambda$.   The peak in $(\w /c)^2 \la |R^>(q,\w )|^2\ra$ for $ q < 
- (\w /c)$ arises from the terms in the expansion of $R^>(q,\w )$ of 
first and third orders in $\zx$; that for $q > (\w /c)$ arises from 
the second-order term.  We see from this figure that $\la |R^>(q,\w 
)|^2\ra$ indeed vanishes for $-(\w /c) < q < (\w /c)$.  With our 
choice of $\Delta k = 0.3 (k_1(\w )-(\w /c))$, a calculation of 
$R^>(q,\w)$ to, say, fourth order in $\zx$ would have yielded a small 
nonzero contribution to $\la |R^>(q,\w )|^2\ra$ in the  radiative 
region $-(\w /c) < q < (\w /c)$, but even that could be suppressed by 
choosing $\Delta k < 0.25 (k_1(\w )-(\w /c))$.

In Fig.~4 we show that the same result is obtained if we increase the
roughness of the surface by increasing $\delta$ from $10$~nm to
$30$~nm, keeping the same values for the remaining material and
roughness parameters that were used in obtaining the results plotted
in Fig.~3.  In this case a computer simulation approach based on the
numerical solution of Eq.~(2.7) was used. The expression (4.4) was
used for $T(x_1)$, and the results for 50 realizations of the surface
profile function were averaged to obtain the results plotted in this
figure. in Fig.~4b four peaks are easily seen. They correspond to
the real parts of the wavenumbers of the scattered surface plasmon
polaritons resulting from the scattering of an incident surface
plasmon polariton of wavenumber
$k(\omega)=k_1(\omega)+ik_2(\omega)=(1.0741+i 0.0026)\omega/c$.
Earlier in this section we predicted that the real parts of these
wavenumbers should be in the vicinity of $\pm k_1(\omega)$, $\pm 3
k_1(\omega)$, etc.  These predictions fit very well with what can be
observed from Fig.~4b, even though the peaks at $\pm 5 k_1(\omega)$,
and higher, cannot be distinguished from the noisy background due to
the low number of samples on which this figure is based. It should be
observed that the peak at $k_1(\omega)$ is stronger then the one at
$3k_1(\omega)$, even though the latter is of first order (in
perturbation theory) while the former is of second order. We believe
that this is related to the fact that both the surface plasmon
polaritons resulting from a single interaction with the rough surface
can be scattered into surface polaritons of (real) wavenumber
$k_1(\omega)$, which in this sense gives it a multiplicity of two.
The surface polaritons at $3k_1(\omega)$ can, in addition to the single
scattering contribution, only be reached by scattering processes of
order three or higher, which are all weak for the roughness assumed here.

Thus,  the method presented in this work provides an
effective way of producing random surfaces that suppress leakage.

\setcounter{equation}{0}
\section{Conclusions}

In this paper we have presented an approach to generating a
one-dimensional random surface that suppresses leakage.  We have shown
by perturbative and numerical simulation calculations that surfaces
generated in this way indeed possess this property.

\section*{Acknowledgments}

The research of A.A.M. and T. A. Leskova was supported in part by Army
research Office Grant DAAG 55-98-C-0034.  I. S. would like to thank
the Research Council of Norway (Contract No. 32690/213) and Norsk
Hydro ASA for financial support.  The work of E. R. M. was supported
by CONACYT Grant 3804P-A.  This work has also received support from
the Research Council of Norway (Program for Supercomputing) through a
grant of computer time.


\begin{references}
\bibitem{1}
  D. Sornette, L. Macon, and J. Coste, J. Phys. (France) {\bf 
49}, 1683 (1988).
\bibitem{2} 
 J.-P. Desideri, L. Macon, and D. Sornette, Phys. Rev. Lett. 
{\bf 63}, 390 (1989).
\bibitem{3} 
 J. A. S\'anchez-Gil and A. A. Maradudin, Phys. Rev. B{\bf 56}, 
1103 (1997).
\bibitem{4}
  F. Pincemin and J. -J. Greffet, J. Opt. Soc. Am. B{\bf 13}, 1499 (1996).
\bibitem{5
}  Lord Rayleigh, {\it The Theory of Sound}, 2nd ed. (MacMillan, 
London, 1896), Vol. II,  pp. 89-96, and 297-311.
\bibitem{6}
  M. F. Pascual, W. Zierau, T. A. Leskova, and A. A. Maradudin, 
Opt. Commun. {\bf 155}, 351 (1998).
\end{references}
\end{document}